%% file: mm-dinov2.tex
\documentclass[runningheads]{llncs}
\usepackage[T1]{fontenc}
\usepackage{graphicx,verbatim}
\usepackage{breakcites}

\usepackage{booktabs}
\usepackage{multirow}
\usepackage{amssymb} % For \checkmark
\usepackage{pifont} % For \ding{55} (×)

\usepackage{xcolor} % to define colors by RGB
\usepackage[colorlinks=true,
            linkcolor=customorange,
            citecolor=customorange,
            urlcolor=customgreen,
            pdfborder={0 0 0}]
           {hyperref}

\definecolor{customorange}{RGB}{247,120,30}
\definecolor{customgreen}{RGB}{34,139,34} % forest green, fitting contrast to orange

\begin{document}
\title{
    MM-DINOv2: Adapting Foundation Models for Multi-Modal Medical Image Analysis
    % : A Framework for Robust Glioma Classification
    % Leveraging DINOv2 for Multi-Modal MRI classification.
}
\titlerunning{Multi-Modal DINOv2 for Medical Image Analysis}
%

% \begin{comment}

\author{%
    Daniel Scholz\inst{1,2,3} \and
    Ayhan Can Erdur\inst{3,4}\and
    Viktoria Ehm\inst{2,5} \and
    Anke Meyer-Baese\inst{6,9} \and
    Jan C Peeken\inst{4,7,8} \and
    Daniel Rueckert\inst{2,3,*} \and
    Benedikt Wiestler\inst{1,2,*}
}

% index{Scholz, Daniel}
% index{Erdur, Ayhan Can}
% index{Ehm, Viktoria}
% index{Meyer-Baese, Anke}
% index{Peeken, Jan C.}
% index{Wiestler, Benedikt}
% index{Rueckert, Daniel}

\authorrunning{D. Scholz et al.}



\institute{%
    Chair for AI for Image-Guided Diagnosis and Therapy, Technical University of Munich (TUM) and TUM University Hospital, Munich, Germany \and
    Munich Center for Machine Learning (MCML), Munich, Germany \and
    Chair for AI in Healthcare and Medicine, Technical University of Munich (TUM) and TUM University Hospital, Munich, Germany \and
    Department of Radiation Oncology, TUM University Hospital, Munich, Germany \and
    Chair for Computer Vision and Artificial Intelligence, Technical University of Munich (TUM), Munich, Germany \and
    Department of Scientific Computing, Florida State University, Tallahassee, FL, USA \and
    Deutsches Konsortium für Translationale Krebsforschung (DKTK), Partner Site Munich, Munich, Germany \and
    Institute of Radiation Medicine (IRM), Department of Radiation Sciences (DRS), Helmholtz Center Munich, Munich, Germany \and
    Institute for Advanced Study, Technical University of Munich (TUM), Munich, Germany
    \\
    $^\mathrm{*}$contributed equally as senior authors \\
    \email{daniel.scholz@mri.tum.de}
}

% \end{comment}

\maketitle              % typeset the header of the contribution

\setcounter{footnote}{0}

\input{0_abstract}
\input{1_introduction}
\input{2_related_work}
\input{3_method}
\input{4_results}
\input{5_discussion}

% \begin{comment}  %% removed for anonymized MICCAI 2025 submission.

% The following acknowledgement and disclaimer sections should be removed for the double-blind review process.
% If and when your paper is accepted, reinsert the acknowledgement and the disclaimer clause in your final camera-ready version.

\begin{credits}
    \subsubsection{\ackname}
    This study was supported by the DFG, grant \#504320104.
    % A bold run-in heading in small font size at the end of the paper is
    % used for general acknowledgments, for example: This study was funded

    \subsubsection{\discintname}
    The authors have no competing interests to declare that are
    relevant to the content of this article
    % It is now necessary to declare any competing interests or to specifically
    % state that the authors have no competing interests. Please place the
    % statement with a bold run-in heading in small font size beneath the
    % (optional) acknowledgments\footnote{If EquinOCS, our proceedings submission
    % system, is used, then the disclaimer can be provided directly in the system.},
    % for example: The authors have no competing interests to declare that are
    % relevant to the content of this article. Or: Author A has received research
    % grants from Company W. Author B has received a speaker honorarium from
    % Company X and owns stock in Company Y. Author C is a member of committee Z.
\end{credits}

% \end{comment}

% \newpage

\bibliographystyle{splncs04}
\bibliography{bibliography}

\end{document}

%% file: 0_abstract.tex
\begin{abstract}
    Vision foundation models like DINOv2 demonstrate remarkable potential in medical imaging despite their origin in natural image domains.
    However, their design inherently works best for uni-modal image analysis, limiting their effectiveness for multi-modal imaging tasks that are common in many medical fields, such as neurology and oncology.
    While supervised models perform well in this setting, they fail to leverage unlabeled datasets and struggle with missing modalities—a frequent challenge in clinical settings.
    To bridge these gaps, we introduce \textbf{{MM-DINOv2}}, a novel and efficient framework that adapts the pre-trained vision foundation model DINOv2 for multi-modal medical imaging.
    Our approach incorporates multi-modal patch embeddings, enabling vision foundation models to effectively process multi-modal imaging data.
    To address missing modalities, we employ full-modality masking, which encourages the model to learn robust cross-modality relationships.
    Furthermore, we leverage semi-supervised learning to harness large unlabeled datasets, enhancing both the accuracy and reliability of medical predictions.
    Applied to glioma subtype classification from multi-sequence brain MRI, our method achieves a Matthews Correlation Coefficient (MCC) of 0.6 on an external test set, surpassing state-of-the-art supervised approaches by {+11.1}\%.
    Our work establishes a scalable and robust solution for multi-modal medical imaging tasks, leveraging powerful vision foundation models pre-trained on natural images while addressing real-world clinical challenges such as missing data and limited annotations.\footnote{The code is publicly available at: \url{https://github.com/daniel-scholz/mm-dinov2}.}

    \keywords{DINOv2  \and Semi-Supervised Learning \and Multi-modal MRI \and Glioma}

\end{abstract}

%% file: 1_introduction.tex
\section{Introduction}
% (2 pages incl abstract)
Vision foundation models, particularly DINOv2~\cite{oquabDINOv2LearningRobust2023}, have demonstrated significant potential in medical image analysis through radiological benchmarks across modalities like MRI, CT, and X-rays~\cite{perez-garciaExploringScalableMedical2025,muller-franzesMedicalSliceTransformer2024,baharoonEvaluatingGeneralPurpose2024,songGeneralPurposeImage2024}. 
However, existing approaches remain constrained to uni-modal analyses or employ suboptimal multi-modal strategies, such as treating MRI sequences as RGB channels~\cite{huangComparativeAnalysisImageNet2024}. 
This limitation prevents their application to clinical tasks requiring joint interpretation of multiple modalities common in fields like oncology or neurology.

Current supervised models for glioma subtype classification achieve competent results when all four standard MRI sequences (T1w, T1ce, T2w, FLAIR) are available~\cite{foltyn-dumitruImpactSignalIntensity2023,vandervoortCombinedMolecularSubtyping2023,scholzImbalanceawareLossFunctions2024}.
However, these approaches do not utilize large-scale unlabeled datasets like BraTS~\cite{baidRSNAASNRMICCAIBraTS20212021}, which contains over 2,000 multi-institutional MRI scans originally curated for segmentation tasks. 
Further, real-world clinical data often suffers from missing modalities — 15\% of patients in routine practice lack at least one essential MRI sequence due to acquisition constraints, protocol variations, or artifacts~\cite{pemberton2023multi}. 
Existing methods fail to address this variability, as they either rigidly require fixed modality inputs or process sequences in isolation.

This work addresses these limitations by proposing a novel framework to adapt pre-trained vision foundation models to the specific requirements of multi-modal medical image analysis. 
Our approach leverages both labeled and unlabeled data while being robust to missing sequences. The contributions of this work are as follows:

\begin{enumerate}
    \item We propose a novel approach to \textbf{adapt pre-trained vision transformers} such as DINOv2 for medical imaging tasks, including a \textbf{new multi-modal patch embedding} tailored for \textbf{multi-modal imaging data}.
    \item We present an adaptive vision transformer architecture that can \textbf{handle missing sequences} during training and evaluation. To this end, we extend the existing masking objective with \textbf{full modality masking} to encourage the model to learn cross-modality relations.
    \item We demonstrate \textbf{improved glioma subtype classification} by effectively utilizing large amounts of unlabeled data through semi-supervised learning. 
\end{enumerate}

%% file: 2_related_work.tex
% \section{Related Work}
% \subsection{Deep learning in glioma subtype classification}
% \begin{itemize}
%     \item supervised focused on class imbalance: \cite{scholzImbalanceawareLossFunctions2024}
%     \item joint segmentation and classification: \cite{vandervoortCombinedMolecularSubtyping2023}
%     \item \cite{cluceruImprovingNoninvasiveClassification2022} compare hierarchical classification approach inspired by genetic markers with multi-class classification.
%     \item leave unlabeled data unused.
%     \item semi-supervised \cite{geDeepSemisupervisedLearning2020}: they do not utilize pre-trained vision foundation models, require generative modeling to fill in missing sequences.
% \end{itemize}
% \subsection{Multi-modal DINOv2 Integrations}
% \begin{itemize}
%     \item medical imaging \begin{itemize}
%         \item project maira \cite{bannurMAIRA2GroundedRadiology2024} with text
%         \item DINO RGB \cite{huangComparativeAnalysisImageNet2024} by naive stacking sequences as channels
%     \end{itemize}
%     \item vision text: \cite{maniparambilUnimodalMultimodalScaling2024,jiangCLIPDINOVisual2024}
%     \item we are the first to extend dinov2 to multiple imaging modalities, which is common in medical imaging.
% \end{itemize}

\section{Related Work}

\subsubsection*{Deep Learning in Glioma Subtype Classification}
Glioma subtype classification is crucial for prognosis and treatment planning and has been tackled with deep learning-based approaches in many works.
Van der Voort et al.~\cite{vandervoortCombinedMolecularSubtyping2023} combine segmentation and classification tasks to improve glioma subtype classification.  
Cluceru et al.~\cite{cluceruImprovingNoninvasiveClassification2022} compare hierarchical classification inspired by genetic markers with standard multi-class classification approaches.  
The class imbalance in glioma is addressed in~\cite{scholzImbalanceawareLossFunctions2024}, where they utilize imbalance-aware supervised loss functions.  
Ge et al.~\cite{geDeepSemisupervisedLearning2020} propose a semi-supervised framework that relies on generative models to impute missing sequences.
While these works tackle the glioma subtype classification problem, they do not yet make use of the powerful available foundational models.

\subsubsection*{Multi-Modal DINOv2 Integrations}
Foundational models such as DINOv2 have demonstrated strong representation learning in multi-modal settings.
\cite{maniparambilUnimodalMultimodalScaling2024,jiangCLIPDINOVisual2024} focus vision-text integrations but do not address multi-modal imaging.
Further,  \cite{bannurMAIRA2GroundedRadiology2024} integrates DINOv2 with text data in a medical context for radiology applications.
So far, multi-modal imaging with DINOv2 has only been addressed in~\cite{huangComparativeAnalysisImageNet2024}, where they adapt DINOv2 for medical imaging by na\"ively stacking modalities as RGB channels, limiting its effectiveness for multi-modal data.  
Our work is the first to extend DINOv2 to handle multiple imaging modalities, a common requirement in medical imaging while addressing robustness to missing modalities and leveraging semi-supervised learning.

%% file: 3_method.tex
\section{Materials and Methods}
\begin{figure}[th]
    \centering
    \includegraphics[width=1.02\linewidth]{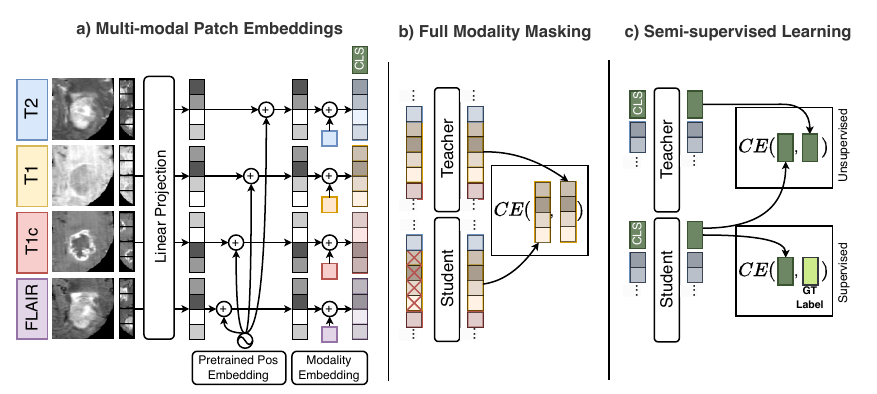}
    \caption{\textbf{Schematic representation of our proposed adaptations to DINOv2.} To leverage the rich synergies of multi-modal imaging data, we (a) define modality-wise positional and individual modality embeddings (Sec. \ref{mmpe}), (b) introduce full modality masking to improve robustness against missing sequences (Sec. \ref{mmr}), and (c) leverage existing labels in a semi-supervised setup (Sec \ref{sse}).}
    \label{fig:pipeline}
\end{figure}

% (2.5 pages)
Our goal is to enable the use of the pre-trained vision foundation model DINOv2 for multi-modal medical imaging tasks. % as it shows a flexible and optimized pre-training paradigm.
%  has already shown great promise in medical imaging~\cite{huangComparativeAnalysisImageNet2024,bannurMAIRA2GroundedRadiology2024}.
To achieve this, we introduce substantial modifications to three key components of DINOv2: the patch embeddings in the vision transformer (ViT)~\cite{dosovitskiyImageWorth16x162020} backbone, the masked image modeling objective, and the image-level objective.
% These changes are specifically designed to address challenges unique to multi-modal medical data, such as the need to process multiple imaging sequences, robustness to missing modalities, and effective utilization of partially available labels.
An overview of our adaptations is shown in Figure~\ref{fig:pipeline}.

\subsection{Multi-modal Patch Embeddings}\label{mmpe}
ViTs, the backbones of DINOv2, treat an image as a sequence of patches, which are flattened and projected into a sequence of patch embeddings $\mathbf{z}_p, \mathrm{with}~p\in \mathcal{P}$, where $\mathcal{P}$ is the set of patches.
The ViTs rely on positional embeddings to encode the order and, hence, the spatial relationships between patches.
However, these learned embeddings are designed for uni-modal data, treating all input patches as originating from a single image.
Applying this directly to multi-modal images discards the valuable multi-modal information about each input patch.
Furthermore, pre-trained vision foundation models lack mechanisms to distinguish input imaging modalities, which is crucial for modality-specific feature extraction, as our experiments show.
To address these limitations, we adapt the patch embedding mechanism in the ViT in two ways.
First, we make sure the pre-trained positional embeddings are applied separately to the patches of each modality rather than treating all modalities as a single input image.
This ensures that the spatial relationships within each modality are preserved.
Formally, we denote the positional embeddings as $\{\mathbf{z}_i\}_{i\in \{1,...,|\mathcal{P}|\}}$.
These positional embeddings are learned during the DINOv2 pre-training and continuously optimized during the pre-training.
Second, we introduce modality-specific embeddings.
These embeddings are a learnable set of feature vectors $\{\mathbf{z}_m\}_{m\in \mathcal{M}}$ with the set of modalities $\mathcal{M}$.
These vectors are initialized as $\mathbf{z}_m \sim \mathcal{N}(\mathbf{0},\mathbf{I})$, yielding the patch embedding $\mathbf{z}_{i,m}$ for the $i$-th patch in modality $m$ as: $\mathbf{z}_{i,m} = \mathbf{z}_{p} + \mathbf{z}_{i} + \mathbf{z}_{m}$.
Therefore, the model can efficiently learn modality-specific representations in these sequence embeddings.

% In summary, by applying pre-trained positional embeddings per modality and introducing learnable modality-specific embedding, we enable pre-trained ViTs, and, hence, DINOv2 for multi-modal medical imaging tasks.

\subsection{Missing Modality Robustness}\label{mmr}

In clinical practice, it is common for patients to lack one or more MRI sequences due to acquisition constraints or artifacts.
To ensure robustness to such cases, we extend the patch-level objective used in DINOv2.
The DINOv2 pre-training employs two ViTs, a student and a teacher, which predict feature representations for each input patch.
% The student tries to match the teacher's predictions.
Input patches to the student are randomly masked while the teacher receives unmasked patches to enforce meaningful representations in the predicted patch representations.
A cross-entropy~($CE$) loss on the masked patches is computed, where the teacher's prediction serves as a target for the student.
We leverage this dynamic to encourage robustness to missing sequences by adding full sequence dropout.
We mask all patches corresponding to one modality in the student's input sequence while the teacher network receives unmasked patches.
The loss remains the same as in DINOv2~\cite{oquabDINOv2LearningRobust2023,zhouImageBERTPretraining2021}.
% , defined as:
% $\mathcal{L}_{patch} = - \sum_{i \in I} p_{t,i} \log p_{s,i}$,
% where \( I \) represents the indices of masked patches, \( p_t \) are the teacher's prototypes, and \( p_s \) are the student's prototypes.
This strategy encourages robust feature learning by forcing the student network to predict token representations for masked modalities based on cross-modality relationships learned from available sequences.

\subsection{Semi-supervised Extension}\label{sse}

While DINOv2 is a self-supervised pre-training method, clinical applications often involve datasets with some diagnostic labels available.
To leverage these labels during pre-training, we incorporate a semi-supervised mechanism inspired by \cite{finiSemisupervisedLearningMade2023}. %, which was originally developed for DINO~\cite{caronEmergingPropertiesSelfSupervised2021a}.
Therefore, we extend the image-level objective from DINOv2:
The student and teacher receive differently augmented crops of the same input image.
Both networks produce image-level probability distributions, \textit{prototype scores}, which are non-linear projections of the CLS token passed through a softmax function.
A cross-entropy ($CE$) loss is calculated between the student's and the teacher's prototype scores, with the latter acting as a pseudo-label:
$\mathcal{L}_{image} = - p_{t} \log p_s$, where \( p_t \) and \( p_s \) represent teacher and student prototype scores.
Since this cross-entropy loss is also often used in supervised learning, it can be easily integrated by replacing the teacher's pseudo-labels with real labels when available.
Fini et al.~\cite{finiSemisupervisedLearningMade2023} exploit this by formulating a joint loss between the supervised loss and the image-level objective.
Here, the prototype score's dimensionality must match the dataset's number of classes.
Intuitively, this guides prototype generation so that CLS token clusters align with known class labels.
% In summary, we extend DINOv2 with a semi-supervised mechanism that leverages partially available labels during pre-training.
% This addition allows us to guide representation learning while maintaining scalability across large unlabeled datasets.

\subsection{Training Setup}
For all experiments, we use the ViT-B/14 model pre-trained with DINOv2 as model initialization.
Since the provided checkpoints do not include weights for the patch- and image-level heads, we randomly initialize these heads and train only them for 10 epochs before unfreezing the entire model.
The output size of both heads is set to three to match the number of classes in our dataset.
We train the model for 200 epochs on a single NVIDIA A40 or A100 GPU with a batch size of 64 and 42 steps per epoch.
The base learning rate is set to $1\mathrm{e}{-4}$.
Positional embeddings are interpolated to match the size of our input images.
Global crops are resized to $98 \times 98$ with sizes in the range $(0.5, 1.0)$ of the input image, while local crops are resized to $56 \times 56$ with sizes in the range $(0.2, 0.5)$.
Crops are always centered around one voxel containing tumor tissue.
For the semi-supervised CE loss, we use a loss weight of 2.0, label smoothing of 0.1, and a temperature scaling of 0.1 in the softmax.

% Our implementation is based on PyTorch (\texttt{torch==2.0.1}) following~\cite{baharoonEvaluatingGeneralPurpose2024}.
% , with additional dependencies including \texttt{xformers==0.0.22}, \texttt{torchmetrics==1.0.3}, and \texttt{torchvision==0.15.2}, following.

\subsection{Dataset}

Our dataset comprises preoperative MR images (T1w, T1ce, T2w, FLAIR) from large public datasets of adult patients with newly diagnosed gliomas \cite{menzeMultimodalBrainTumor2015,bakasAdvancingCancerGenome2017,baidRSNAASNRMICCAIBraTS20212021,suterLUMIEREDatasetLongitudinal2022,bakasUniversityPennsylvaniaGlioblastoma2022,calabreseUniversityCaliforniaSan2022,vandervoortErasmusGliomaDatabase2021,gusevREMBRANDTStudyLarge2018,bakasAdvancingCancerGenome2017}.
% namely BraTS2021~\cite{menzeMultimodalBrainTumor2015,bakasAdvancingCancerGenome2017,baidRSNAASNRMICCAIBraTS20212021}, LUMIERE~\cite{suterLUMIEREDatasetLongitudinal2022}, UPENN GBM~\cite{bakasUniversityPennsylvaniaGlioblastoma2022}, Rembrandt~\cite{gusevREMBRANDTStudyLarge2018} UCSF-PDGM~\cite{calabreseUniversityCaliforniaSan2022}, EGD~\cite{vandervoortErasmusGliomaDatabase2021}, and TCGA~\cite{bakasAdvancingCancerGenome2017}.
We hold out the TCGA dataset~\cite{bakasAdvancingCancerGenome2017} for external testing and randomly split the remaining datasets in $70/10/20\%$ for training, validation, and internal testing.
This setup yields 2661 (1162 labeled) subjects for training and validation, 296 for internal, and 214 for external testing.
All images provide all four imaging sequences outlined above, while the labeled images have labels from biomarker testing for \textit{IDH} mutation and 1p/19q status in order to classify samples according to the 2021 WHO classification of brain tumors into (a) \textit{IDH} wildtype glioblastoma, (b) \textit{IDH} mutant and 1p/19q intact astrocytoma and (c) \textit{IDH} mutant and 1p/19q codeleted oligodendroglioma~\cite{wen2021WHOClassification2021}.
The class prevalence in the dataset is 80/10/10\% for the three classes, respectively.

All images are resampled to $1\times1\times1$mm isotropic resolution and rigidly registered to the SRI24 atlas~\cite{rohlfingSRI24MultichannelAtlas2009}.
For training, we randomly sample axial, sagittal, and coronal slices from the volume with at least 500 tumor pixels, crop them to $96\times96$, and evaluate on $96\times96$ axial middle slices of the tumor.

%% file: 4_results.tex
\section{Results}

We rigorously evaluate how adapting pre-trained DINOv2 improves multi-modal medical image classification and enhances robustness to missing modalities.
Furthermore, we perform a detailed ablation study to assess the impact of each design choice of our proposed method, demonstrating how these contributions collectively improve downstream performance for the clinical application of glioma subtype classification.

\subsection{Adapting Vision Foundation Models}

To evaluate the effectiveness of our proposed adaptations, we compare two scenarios: fully supervised training and semi-supervised pre-training, followed by linear evaluation.
Both scenarios include the following two architectures: (1) na\"ive DINOv2 with concatenated T1ce, T2w, and FLAIR sequences as RGB inputs~\cite{huangComparativeAnalysisImageNet2024}
and (2) our multi-modal adaptation of DINOv2 (MM-DINOv2).
Table~\ref{tab:main-table} summarizes the results across all scenarios.
Given the imbalanced and multi-class nature of the classification task, we employ the Matthews Correlation Coefficient (MCC) as the primary evaluation metric~\cite{maier-heinMetricsReloadedRecommendations2024}, alongside AUROC for classifier calibration and class-wise F1 scores to assess per-class performance.

\paragraph{Supervised Results}
In the fully supervised setting, only labeled data are used (approximately 50\% of the full dataset).
The DINOv2 variants are fully fine-tuned except for the ``frozen'' row in Table~\ref{tab:main-table}, which only utilizes the pre-trained weights.
ResNet34~\cite{heDeepResidualLearning2015} serves as a strong baseline due to its robustness in low-data regimes.
It outperforms both linear and fully fine-tuned DINOv2 with na\"ively stacked MRI sequences as in the RGB channels model.
This result highlights the importance of our architectural adaptations for multi-modal data, including the drawbacks of neglecting one MRI input sequence.
Our adapted multi-modal DINOv2 with full fine-tuning achieves superior results, indicating the positive influence of our design choices for multi-modal adaptation.

\paragraph{Semi-Supervised Results}
In the semi-supervised setting, both labeled and unlabeled data are utilized.
We add the proposed semi-supervised extension to all used DINOv2 variants and find our MM-DINOv2 outperforming the RGB DINOv2.
Our adapted multi-modal DINOv2 further improves for two out of three classes compared to the fully fine-tuned model, highlighting the strength of incorporating unlabeled images into the training process.
Nevertheless, we assume that a large amount of unlabeled data introduces more class imbalance compared to the supervised setting, such that the challenging and underrepresented class of oligodendroglioma decreases in performance.

%---------------main table--------------------

\begin{table}[th]
    \centering
    \caption{\textbf{Comparison of supervised and semi-supervised approaches for glioma subtype classification} using Matthews Correlation Coefficient (MCC), AUROC, and class-wise F1 scores. Results are reported for RGB DINOv2 with concatenated modalities as RGB channels and our adapted multi-modal DINOv2 (MM-DINOv2) with continuous pre-training. MM-DINOv2 outperforms the comparison methods in all metrics (\textbf{best}, \underline{best in section}).}\label{tab:main-table}
    \centering
    \begin{tabular}{lcccccccccc}
        \toprule
        Method                                                                                        & \multicolumn{2}{c}{MCC} & \multicolumn{2}{c}{AUROC} & \multicolumn{6}{c}{F1 Score}                                                                                                                                                         \\ \cmidrule{2-11}
                                                                                                      &                         &                           &                              &                  & \multicolumn{2}{c}{Astro} & \multicolumn{2}{c}{GBM} & \multicolumn{2}{c}{Oligo}                                                    \\ \cmidrule{6-11}
                                                                                                      & Int.                    & Ext.                      & Int.                         & Ext.             & Int.                      & Ext.                    & Int.                      & Ext.          & Int.             & Ext.          \\ \midrule
        \multicolumn{11}{l}{\textbf{Supervised}}                                                                                                                                                                                                                                                                                                   \\
        % ResNet34 & 0.61 & 0.51 & 0.94 & 0.74 & -1 & -1 & -1 & -1 & -1 & -1   \\
        ResNet34~\cite{heDeepResidualLearning2015}                                                    & 0.58                    & 0.54                      & 0.92                         & 0.79             & 0.60                      & 0.66                    & 0.92                      & 0.88          & 0.43             & 0.35          \\
        RGB DINOv2 (frozen)~\cite{huangComparativeAnalysisImageNet2024,oquabDINOv2LearningRobust2023} & 0.40                    & 0.27                      & 0.87                         & 0.74             & 0.46                      & 0.31                    & 0.90                      & 0.81          & 0.35             & 0.15          \\
        % MRI-DINO FFT (from scratch) & -1 & -1 & -1 & -1 & -1 & -1 & -1 & -1 & -1 & -1   \\
        RGB DINOv2~\cite{huangComparativeAnalysisImageNet2024}                                        & 0.55                    & 0.52                      & 0.92                         & 0.80             & 0.55                      & 0.64                    & 0.90                      & 0.85          & 0.50             & \textbf{0.37} \\
        MM-DINOv2 (\textbf{ours})                                                                     & \underline{0.68}        & \textbf{0.60}             & \textbf{0.95}                & \textbf{0.89}    & \underline{0.68}          & \textbf{0.71}           & \underline{0.94}          & \textbf{0.89} & \underline{0.62} & 0.33          \\ \midrule
        % \multicolumn{11}{l}{\textbf{Self-supervised}}                                                                                                                                                                                                            \\
        % RGB DINOv2~\cite{huangComparativeAnalysisImageNet2024} & -1 & -1 & -1 & -1 & -1 & -1 & -1 & -1 & -1 & -1   \\
        % % MRI-DINO SSL (from scratch) & -1 & -1 & -1 & -1 & -1 & -1 & -1 & -1 & -1 & -1   \\
        % MM-DINOv2 (\textbf{ours}) & -1 & -1 & -1 & -1 & -1 & -1 & -1 & -1 & -1 & -1   \\ \midrule
        \multicolumn{11}{l}{\textbf{Semi-supervised}}                                                                                                                                                                                                                                                                                              \\
        RGB DINOv2~\cite{huangComparativeAnalysisImageNet2024}                                        & 0.47                    & 0.37                      & 0.84                         & 0.77             & 0.44                      & 0.42                    & 0.93                      & 0.83          & 0.40             & \textbf{0.37} \\
        % MRI-DINO SemiSL (from scratch) & -1 & -1 & -1 & -1 & -1 & -1 & -1 & -1 & -1 & -1   \\
        MM-DINOv2 (\textbf{ours})                                                                     & \textbf{0.74}           & \underline{0.57}          & \textbf{0.95}                & \underline{0.86} & \textbf{0.76}             & \textbf{0.71}           & \textbf{0.96}             & \textbf{0.89} & \textbf{0.67}    & 0.21          \\ \bottomrule
    \end{tabular}
\end{table}

\subsection{Missing Sequence Robustness}

To evaluate the effectiveness of our full modality masking strategy, we compare models trained with and without this design choice on our test sets where one MRI sequence is randomly masked.
Across all metrics, including MCC, AUROC, and class-wise F1 scores, the model trained with full sequence masking consistently outperforms the model trained without it.
Solely, the external test set performance on the majority class (glioblastoma) suffers slightly in terms of F1 score.
These results demonstrate that our masking strategy effectively encourages the model to learn cross-modality relationships, enabling robust performance when one modality is missing during inference.

\begin{table}[th]
    \caption{\textbf{Missing sequence robustness analysis.} We compare two models trained with all four sequences and either with and without full sequence dropout by evaluating their performance with one sequence randomly missing.}
    \label{tab:full-sequence-masking}
    \centering
    \begin{tabular}{@{}lcccccccccc@{}}
        \toprule
        Full Sequence Masking & \multicolumn{2}{c}{MCC} & \multicolumn{2}{c}{AUROC} & \multicolumn{6}{c}{F1 Score}                                                                                                                                                   \\ \midrule
                              &                         &                           &                              &               & \multicolumn{2}{c}{Astro} & \multicolumn{2}{c}{GBM} & \multicolumn{2}{c}{Oligo}                                                 \\ \cmidrule(l){6-11}
                              & Int.                    & Ext.                      & Int.                         & Ext.          & Int.                      & Ext.                    & Int.                      & Ext.          & Int.          & Ext.          \\ \midrule
        No                    & 0.46                    & 0.41                      & 0.86                         & 0.82          & 0.47                      & \textbf{0.58}           & 0.92                      & 0.83          & 0.36          & 0.13          \\
        Yes                   & \textbf{0.57}           & \textbf{0.46}             & \textbf{0.89}                & \textbf{0.83} & \textbf{0.61}             & 0.56                    & \textbf{0.94}             & \textbf{0.86} & \textbf{0.42} & \textbf{0.28} \\\bottomrule
    \end{tabular}
\end{table}

\subsection{Ablation Study}

We conduct an ablation study to rigorously evaluate the impact of each design choice in our model, ranging from simply concatenating all tokens from all modalities to our full multi-modal adaptation.
The results, shown in Table~\ref{tab:ablation-study}, are evaluated in the semi-supervised setting, as it achieved the best overall performance.
We observe poor performance initially with a single global positional embedding for all tokens from all modalities treated as a single input modality, which corresponds to spatially concatenating modalities.
This is expected, as spatial concatenation implies false spatial correlations between modalities.
Subsequent design choices, including modality-specific embeddings and  full sequence masking, consistently improve performance.
%The only exception is the minority class \textit{oligodendroglioma}, where performance remains suboptimal.
%This result is consistent across most baselines and reflects the inherent difficulty of capturing features for this underrepresented class which might be exacerbated by the additional unlabeled data.

\begin{table}[th]
    \centering
    \caption{\textbf{Ablation study over our design choices.} Adding our proposed adaptations continuously improves the glioma subtype classification performance.}\label{tab:ablation-study}
    \begin{tabular}{@{}lcccccccccc@{}}
        \toprule
        Semi-supervised           & \multicolumn{2}{c}{MCC} & \multicolumn{2}{c}{AUC} & \multicolumn{6}{c}{F1 Score}                                                                                                                                                   \\ \cmidrule(l){2-11}
                                  &                         &                         &                              &               & \multicolumn{2}{c}{Astro} & \multicolumn{2}{c}{GBM} & \multicolumn{2}{c}{Oligo}                                                 \\ \cmidrule(l){6-11}
                                  & Int.                    & Ext.                    & Int.                         & Ext.          & Int.                      & Ext.                    & Int.                      & Ext.          & Int.          & Ext.          \\ \midrule
        Concat Tokens             & 0.40                    & 0.29                    & 0.84                         & 0.73          & 0.41                      & 0.31                    & 0.93                      & 0.82          & 0.29          & \textbf{0.22} \\
        + Per-Image Pos Embedding & 0.63                    & 0.36                    & 0.93                         & 0.79          & 0.68                      & 0.45                    & 0.95                      & 0.83          & 0.35          & 0.12          \\
        + MRI Sequence Embedding  & \textbf{0.74}           & 0.49                    & 0.94                         & \textbf{0.86} & 0.71                      & 0.65                    & \textbf{0.97}             & 0.86          & \textbf{0.71} & 0.13          \\
        + Full Sequence Masking   & \textbf{0.74}           & \textbf{0.57}           & \textbf{0.95}                & \textbf{0.86} & \textbf{0.76}             & \textbf{0.71}           & 0.96                      & \textbf{0.89} & 0.67          & 0.21          \\ \bottomrule
    \end{tabular}
\end{table}

%% file: 5_discussion.tex
\section{Conclusion}
%SUGGESTION BENE
This work introduces a novel adaptation strategy for vision foundation models like DINOv2 to effectively address multi-modal medical imaging challenges.
Our approach enhances robustness to missing modalities and leverages partially available labels to improve performance on clinical tasks such as glioma subtype classification.
Our findings highlight the importance of incorporating semi-supervised learning, which outperformed supervised training by utilizing all available data.
Moreover, we demonstrated that methodological adaptations tailored to multi-modal medical image analysis are essential for tasks integrating multiple imaging modalities.
Notably, our full sequence masking strategy effectively addressed the challenge of missing MRI sequences, a frequent issue in clinical workflows.
Our contributions complement existing approaches like Rad-DINO~\cite{perez-garciaExploringScalableMedical2025} while emphasizing multi-modal data integration and robustness, critical factors for real-world clinical scenarios.
Furthermore, our results align with prior studies on semi-supervised learning in medical imaging~\cite{geDeepSemisupervisedLearning2020} while extending these insights to foundation models.
Future directions could explore combining our approach with methods like the Medical Slice Transformer~\cite{muller-franzesMedicalSliceTransformer2024} to extend foundation models to 3D volumetric imaging tasks, broadening their applicability to radiology and beyond.
By addressing critical challenges in multi-modal medical imaging, we hope our work inspires further innovation in diagnostic tools and contributes to improved patient care.